\documentclass[lettersize,journal]{IEEEtran}

\usepackage{url}
\usepackage{times}
\usepackage{amsmath}
\usepackage{epsfig}
\usepackage{multirow}
\usepackage{longtable}
\usepackage{amsfonts}
\usepackage{amssymb}%
\usepackage{color}
\usepackage{bm}
\usepackage{epstopdf}
\usepackage{subfigure}
\usepackage{amsfonts}
\usepackage{cite}
\usepackage{graphicx}
\usepackage{amsmath}
\usepackage{times}
\usepackage{latexsym}
\usepackage{bm}
\usepackage{amssymb}
\usepackage[center]{caption2}
\usepackage{stfloats}
\usepackage{cases}
\usepackage{array}
\usepackage{setspace}
\usepackage{fancyhdr}
\usepackage{balance}
        {\par\noindent{\bf Proof.}}%
        {{\hfill{\large$\Box$}}\smallskip}

\DeclareSymbolFont{lettersA}{U}{txmia}{b}{it}
\DeclareMathSymbol{\uptheta}{\mathord}{lettersA}{18}

\begin{document}

\title{
Learning with Dynamics: Autonomous Regulation of UAV Based Communication Networks with Dynamic UAV Crew}

\author{\IEEEauthorblockN{
Ran Zhang,
Bowei Li, 
Liyuan Zhang, 
Jiang (Linda) Xie, 
and Miao Wang}
% \\
% \IEEEauthorblockA{\IEEEauthorrefmark{1}Department of Electrical and Computer Engineering, University of North Carolina at Charlotte, NC, USA,}\\
% \IEEEauthorblockA{\IEEEauthorrefmark{2}}Department of Electrical and Computer Engineering, Carnegie Mellon University, Pittsburgh, PA, USA\\
% \IEEEauthorblockA{\IEEEauthorrefmark{3}}Department of Sensor Engineering, PassiveLogic Inc., Salt Lake City, UT, USA,\\
% \IEEEauthorblockA{\IEEEauthorrefmark{4}}\\
% \IEEEauthorblockA{\IEEEauthorrefmark{5}}School of Telecommunication Engineering, Xidian University, Xi’an, China

}

%
%\thanks{This work was presented in part at the IEEE Globecom 2013~\cite{Miao.GC13}.

\maketitle
\IEEEpeerreviewmaketitle

\begin{abstract}
Unmanned Aerial Vehicle (UAV) based communication networks (UCNs) are a key component in future mobile networking. To handle the dynamic environments in UCNs, reinforcement learning (RL) has been a promising solution attributed to its strong capability of adaptive decision-making free of the environment models. 
%To handle the dynamic environment in UCN-related problems, reinforcement learning (RL) has been a promising solution. Free of environment models, RL is strongly capable to make adaptive decisions to the environment dynamics. 
However, most existing RL-based research focus on control strategy design assuming a fixed set of UAVs. Few works have investigated how UCNs should be adaptively regulated when the serving UAVs change dynamically. This article discusses RL-based strategy design for adaptive UCN regulation given a dynamic UAV set, addressing both reactive strategies in general UCNs and proactive strategies in solar-powered UCNs. An overview of the UCN and the RL framework is first provided. Potential research directions with key challenges and possible solutions are then elaborated. Some of our recent works are presented as case studies to inspire innovative ways to handle dynamic UAV crew with different RL algorithms.
\end{abstract}



% \fi
\IEEEpeerreviewmaketitle

%===============================================================================================
\section{Introduction}\label{sec.Intro}
%===============================================================================================
Unmanned aerial vehicles (UAVs) have been demonstrating dazzling potentials in next generation wireless communication and networking. UAVs equipped with wireless transceivers can serve as mobile base stations (BSs) and interconnect to form UAV based communication networks (UCNs). %UCNs stand out in providing highly on-demand services by virtue of flexible 3D mobility, better wireless connectivity with higher chance of Line-of-Sight (LoS) links, and much lower deployment and operational cost due to almost infrastructure-free network construction\cite{zeng2016wireless}. 
As reported in \cite{market}, the UAV market is anticipated to reach 166.7 billion USD by 2031. Driven by the booming market, UCNs have been extensively studied in various aspects such as emergency rescue, network coverage enhancement and extension, crowd/traffic surveillance, cached content delivery, and mobile edge computing\cite{c}. 

Conventional approaches to UCN problems typically adopt alternative optimization, heuristic algorithms or statistical analysis\cite{c}. These methodologies are better fits when the network parameters are fixed. In UCNs, parameters such as the network size and topology, connectivity, channel conditions and service demands are usually dynamically changing, making such methods re-executed every time when the parameters are updated. In addition, for works that consider sequential decision making in a slotted time horizon, computing complexity increases exponentially with the number of slots. This significantly challenges the computing power of decision makers in cases of large network scale or long time horizon.

%With the exponentially increasing network scale and heterogeneity in the future, it will be increasingly difficult for conventional approaches to handle the network dynamics.

Recent advances in reinforcement learning (RL) \cite{bai2023towards} have brought promising solutions to UCN problems. By actively interacting with the environment and learning from the interaction experiences, RL agents are strongly capable to make time-sequential decisions in dynamic environments free of the environment models. RL can be either centralized or distributed. Centralized RL features a single agent making decisions for all the UAVs based on the complete network information\cite{luong2021deep}. While potentially yielding a better overall network performance, the computing power of the agent is significantly challenged. Constant communications between the agent and all the UAVs may incur unacceptable communication overhead and latency in large-scale UCNs. On the contrast, distributed RL, i.e., multi-agent RL (MARL), distributes the training load across the UAVs\cite{dai2022multi}. Each UAV acts as an agent and trains its own policy based on the local observations and limited information sharing with other UAVs for coordinated goals. Scalability can be achieved. %With the increasing size of UCNs, distributed learning is becoming a better choice to achieve scablibility.

Nevertheless, existing RL-based studies on UCNs focus mainly on control policy design given a fixed set of UAVs. Few works have dabbled how the network should be adaptively regulated when the serving UAV crew dynamically change. On one hand, the serving UAV crew may passively change at times: UAVs have to quit the network when they run out of batteries; supplementary UAVs can also join the serving crew whenever needed. Fluctuations in the network performance will be inevitably incurred, thus calling for \textit{passively} responsive regulation strategies when such changes happen. On the other hand, the future UAV models are expected to be solar-power rechargeable\cite{morton2015solar}. This leaves the network great initiative to \textit{proactively} control UAVs' quit and join-in rather than performing a passive response strategy. %\footnote{Solar panels on UAVs have satisfying charging rates only at high altitudes\cite{}. Hence we consider that a UAV will temporally quit the network and elevate high into the sky for solar-charging.}. 
When user demand is low, certain UAVs can be scheduled for solar charging even if they are not in bad need. They can be later called back to replace other UAVs or meet increasing demands. 
In this manner, a self-sustainable UCN can be established based on benignant charging-serving cycles, optimizing the target network performance subject to the constraints of network sustainability and time-varying user demand.

%Nevertheless, existing studies on MARL-based UAV networks usually assume a fully cooperative relationship among UAVs which neglects UAVs' own benefits. Yet in many occasions, UAVs may have a hybrid relationship, i.e., cooperate and compete at the same time. One example is in solar-powerd UAV networks, where each UAV may want to maximize its own residual energy while working together to satisfy the network constraints. Another example is when the UAVs belong to different operators. Under the hybrid relationship, simply following the existing design of the MARL algorithms may lead to unsatisfying distributed policies. Novel approaches must be identified or developed to model the cooperative-competitive interaction among the agents and guide the value function updates for each agent during the training stage.   

%\subsection{Proposed Research Goals}

To this end, we discuss in this article how, under the RL framework, UCNs can responsively react to and proactively control the dynamic change of the serving UAV crew in either a centralized or distributed manner. Specifically, the following two aspects are elaborated.
\begin{itemize}
    \item {\em Design of RL-Based Responsive Strategies to Dynamic UAV Crew in General UCNs}. With such strategies, when any UAVs are about to quit or join, the strategy will direct the active UAVs to take autonomous yet synergetic actions to minimize (or maximize) the performance loss (or gain). As the UAVs serve as mobile BSs, the target strategy will feature a joint design of UAV radio resource management (RRM) and trajectory control. In addition, although the strategy does not control the crew change, it is expected to be capable of identifying the upcoming change and actively regulating the UAVs ahead of the change rather than passively reacting after the change. %The proposed research will emphasize designing MARL algorithms to handle dynamic UAV lineup change during the training, developing training mechanisms to promote the agent exploration around the lineup change, and prototyping the envisioned UAV networks to evaluate the timeliness and efficacy of the distributed strategy.
    
    \item {\em Development of RL-Based Proactive Control Strategies on UAV Crew in Solar-Powered Sustainable UCNs}. Unlike the above aspect where the network passively reacts to the change of UAV crew, strategies can be developed to proactively control the change in solar-powered UCNs. Leveraging the spatio-temporal variability of the user service demand, RL-based strategies will be designed to optimize UAVs' charging profiles. The optimization aims to strike the balance between UAVs' individual benefits and overall network performance while fulfilling the network sustainability and time-varying user demand. %To model the hybrid cooperate-compete relationship among UAVs, game theories will be integrated into the MARL framework. Research emphases will include prediction of user time-variabilities, integration of game theories into the policy update, and formulation and solutions to the resultant MARL problems.
\end{itemize}

In the following, an overview is first presented on models of UCNs together with the RL framework. Potential research problems, challenges, and approaches are then discussed, followed by our recent related works as case studies. We expect the research outcomes to provide valuable inspirations and benchmarks to the autonomous management of UCNs under a dynamic network setup.

%===============================================================================================
\section{Overview of UCNs and the RL Framework}\label{sec.2}
\begin{figure}[!ht]
\centering
%\begin{tabular}{l}
%\subfigure[System Model] {\includegraphics[width=2.5in, height=2.2in]{figure_magazine/NetworkModel.eps}} 
%\subfigure[The tailored MARL Framework] {\includegraphics[width=2.5in, height=2.2in]{figure_magazine/MARLFr.eps}}
%\end{tabular}
{\includegraphics[width=2.6in]{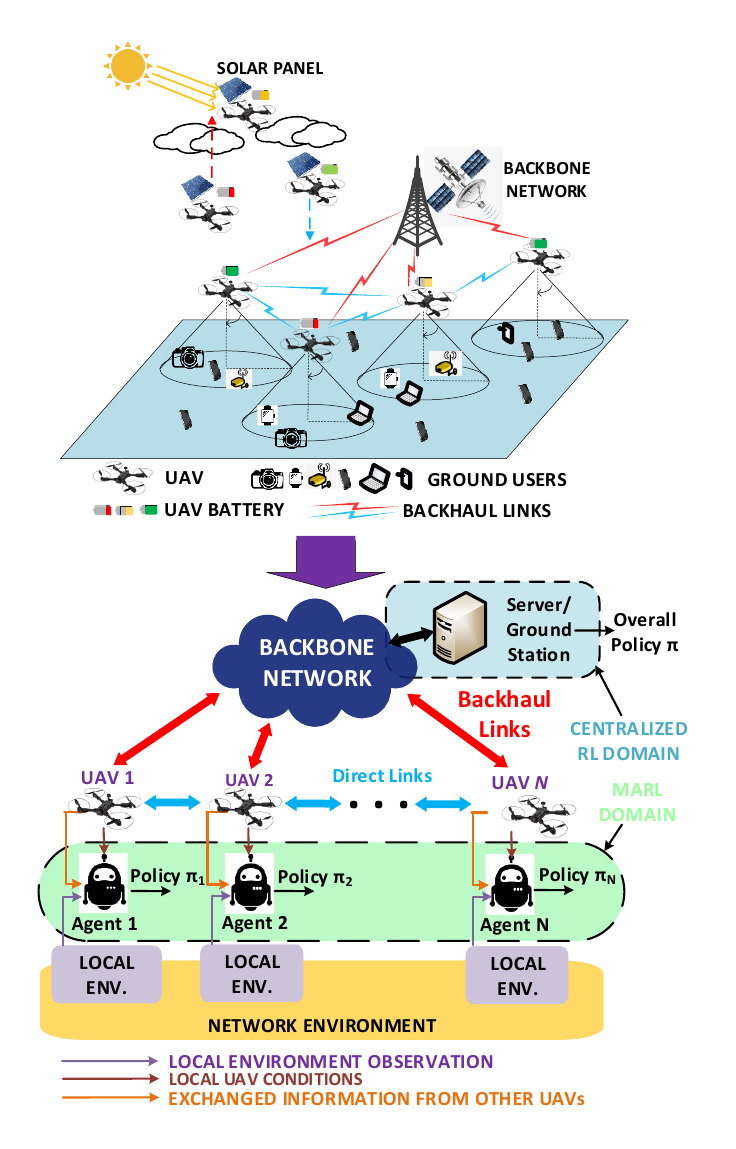}}
\caption{\small{Network model and the underlying RL framework for UCNs.}}\label{fig.systemmodel}
\end{figure}

%===============================================================================================
The article will focus on developing regulation strategies to responsively react to and proactively control the dynamic change in UAV crew under the RL framework. %The strategies shall be capable of adaptively making joint decisions on RRM and trajectory control (or positioning) for the UAVs when the serving lineup dynamically change, and for solar-powered UAVs, formulating the optimal UAV charging plan to proactively control their quit and re-join to maximize the network performance. 
We study a group of interconnected UAVs flying over a target region, providing communication services to ground users of various types as shown in Fig.\ref{fig.systemmodel}. Each UAV concentrates its transmission power within a certain aperture angle, so its coverage mainly depends on the altitude and transmission power. A UAV can communicate with other UAVs within reach via direct links or backbone networks (e.g., ground BSs or satellites) indirectly. The UAV direct links and backhaul links employ disjoint spectrum from the UAV-ground links, thus avoiding mutual interference. The UAVs are battery-constrained. A serving UAV will quit the network when running out of battery, while a new UAV may join any time. Accordingly, the existing UAVs will be reconfigured (in RRM and positions/trajectories) to minimize (or maximize) the performance loss (or gain). 

For solar-powered UAVs, they can elevate to high altitudes (1000m$\sim$10000m) for high-rate solar charging when they can be unoccupied from users and return later when needed. This creates opportunities to optimally determine UAV charging profiles to enhance the network performance while meeting UCN sustainability and user service requirements. We consider that when a UAV needs solar-charging, it quits the network temporarily and elevates high above the clouds. On one hand, the solar radiation is attenuated exponentially with the thickness of clouds between the sun and solar panel\cite{kokhanovsky2004optical}. The intensity falls
to only $\sim 1/10$ after the first 300$m$ down from the upper cloud edge. Since it does not take
long (e.g., one minute or two) for a UAV to move vertically up
through 300 meters, it is reasonable to charge the UAVs just above the clouds. On the other hand, when a UAV is at a high altitude, its communication to the ground users will degrade significantly and higher interference may be incurred due to increased footprint. Therefore, suspending communication to the ground (i.e., temporally quitting the network) at high altitudes will considerably ease RRM of the UCN.
%\vspace{-6mm}

The timing of the entire system is slotted and synchronized. In every slot, each UAV updates its local observations including local environment situations (e.g., number of served users, solar charging intensity), and local UAV conditions (e.g., 3D coordinates and battery residuals). In centralized RL, a centralized server or a ground control station will collect local observations across UAVs, make RRM and movement decisions for each UAV, and return the decisions via backhaul links. In MARL, each UAV implements its own learning agent and iteratively updates its policy based on its (historical) local observations and optionally the shared information from other UAVs. Inter-UAV information sharing may improve the coordinated learning, yet adds communication overhead and latency. The shared information can be integrated into the local learning states, the reward function design, or the value function updating.

%Under the MARL framework, the timing of the entire system is slotted and synchronized. Each UAV implements its own learning agent, as shown in Fig.\ref{fig.systemmodel}(b). In every slot, agents update their policies based on their (historical) local observations and optionally some deliberately shared information from other UAVs. The local observations consist of local environment situations (e.g., user service demand, solar charging intensity), and local UAV conditions (e.g., 3D coordinates and battery residual). The information from other UAVs can expedite the coordinated learning convergence, yet at a cost of extra communication overhead and latency. Such information can be either integrated as part of the local learning states, in the reward function design, or in the value function updating. %When the local states of the agents are discrete and limited, a Q-matrix will suffice to represent the policy $\pi$. When the local state space is huge or continuous, deep neural networks can be exploited to represent the Q learning  

%===============================================================
\section{Responsive Strategy Design to Dynamic UAV Crew in General UCNs}\label{sec.2}
%===============================================================
When UAV crew change, RL-based response strategies are desired to identify the upcoming change and \emph{actively} regulate the RRM and movements of the existing UAVs. By ``actively", we expect the strategy to take actions in advance when the change is about to happen rather than react passively after the change. The goal is to minimize (maximize) the performance loss (gain) during transition incurred by the change. %Our preliminary works\cite{zhang2020srec,zhang2021learning} have partially solved the problem, but only under a centralized deep RL (DRL) framework with fixed UAV altitude and RRM setting. 
The actions need to include both RRM decisions (e.g., user association, spectrum allocation, power control, etc.) and trajectory design for the best regulation performance. It is much more challenging to jointly optimize multi-UAV RRM and trajectory control under the RL framework given a dynamic UAV crew. We identify the following research directions with key challenges, open issues and potential approaches. 

%-------------------------------------------------------
\subsection{Design of Key Elements in RL}\label{sub.design_Elements}
Dynamic changes in UAV crew impose higher requirements on designing the state space, action space and reward function of the learning agent. The state space needs to be elastic in dimension to accommodate the change, meanwhile being information-inclusive to help identify the upcoming change. In addition, when RRM is configurable, RRM decisions become part of the actions. However, the action space of user association and bandwidth allocation is discrete, while that of power control and UAV positioning is continuous. None of the classical RL algorithms can handle such a mixed action space. %\cite{hu20201}. 
%Furthermore, in an MARL setting, to achieve a shared optimization objective, the design of state space and reward function for individual UAVs may need to integrate useful information from other UAVs. Yet what information to exchange among UAVs remains an open issue due to the tradeoff among convergence performance, complexity and communication overhead. 

%In this task, we plan to study the appropriate design of the state space, action space, and reward function of the MARL algorithm to accommodate the dynamic change of UAV lineup. 
For state space, in addition to the network performance metrics and UAV conditions, the ``in/away" status of all the potentially involved UAVs should also be incorporated to help identify whether a change in the instantaneous reward is due to the change of crew or UAV positions. 
For the ``in/away" status, continuous variables such as battery residual and the join-in countdown timer are preferred over binary indicators so that agent(s) can foresee the upcoming change and make proper regulations ahead of the change. Moreover, agent(s) should maintain a ``maximum" state space to accommodate the maximum possible number of UAVs, but only activate the dimensions of the UAVs whose status are ``in" or to be changed during the learning.

To handle the mixed action space, the advantages of actor-critic RL (AC-RL) and deep Q-learning (DQL) can be potentially combined. AC-RL excels in continuous action space, e.g. power control and positioning, %\cite{yuan2020energy}
%\cite{wang2019continuous}
while DQL is effective for discrete action space, e.g., UAV RRM. %\cite{koushik2019deep}
One approach is to separate the policy into discrete and continuous components. During the learning, two neural networks can be trained: a critic network for the Q-value function and an actor network for the continuous policy. In each iteration, the discrete policy will be first derived from the updated critic network, and then together with the Q-values be fed into the actor network to determine the continuous policy. We expect this two-stage method would address the challenge of hybrid action space.

\subsection{Algorithm Design with Promoted Exploration}\label{subsec.1_promote}
%-------------------------------------------------------

During training, sufficient exploration in state-action space around the timing of UAV crew changes is crucial to optimize the network response. A straightforward way is to increase random \textit{exploration} around the time of change. This can be achieved by increasing the probability of selecting a random action in (deep) Q-learning, or magnifying the random noise to the output action of the actor network in AC-RL. However, our early simulations showed that such methods failed to prompt ``active" reconfiguration ahead of the change or reactive adaptation even after the change. Innovative methods are desired to fully explore the state-action space around the time of change to determine if an ``active" reconfiguration ahead of the change is beneficial.

\begin{figure}[!ht]
  \centering
  \includegraphics[width=3.0in, height=1.4in]{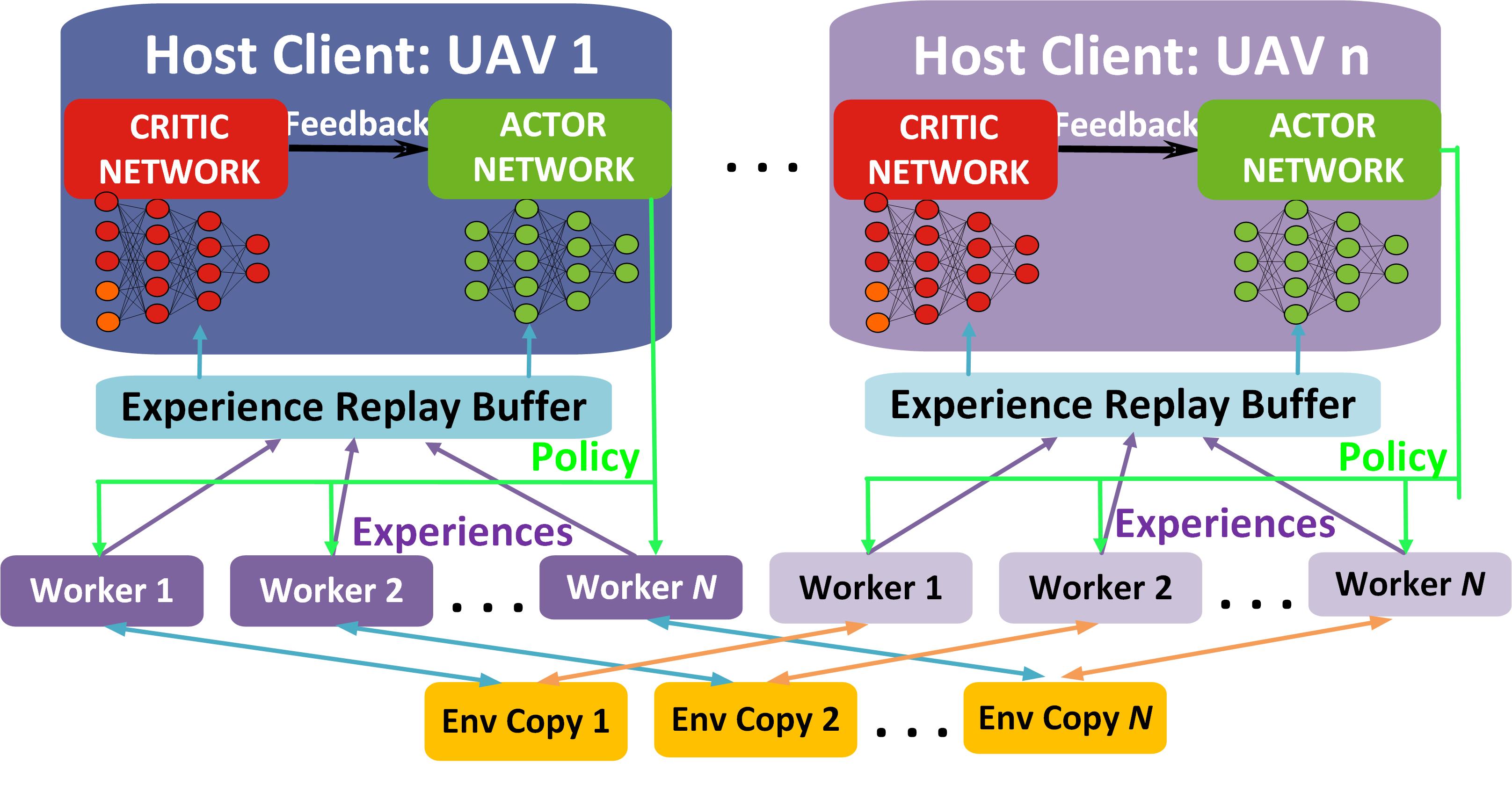}\\
  \caption{Asynchronous parallel computing (APC) diagram.}\label{fig.APC}
\end{figure}

To this end, an alternative to simply increasing the randomness is to reduce the correlation among the collected experiences used for training the agent neural networks, which can be crucial in early training stages. Experiences from one episode are more or less correlated, which may degrade the neural network training in early stages and eventually affect the final training performance. To address this, an asynchronous parallel computing (APC) structure can be exploited\cite{zhang2021learning}, inspired by the asynchronous advantage actor critic (A3C) algorithm\cite{yu2022energy}. 
As shown in Fig.\ref{fig.APC}, in APC, 
%APC applies to actor-critic learning. %when both critic and actor networks are used\cite{mnih2015human}. 
%it considers multiple independent environment copies. 
one agent is composed of a host client and multiple parallel workers, each interacting with an independent environment copy. Such copies are exactly the same but the randomness. The host client maintains unified critic and actor networks and updates both networks using the collected experiences from all workers. Each parallel worker executes the up-to-date policies in its own copy of environment. Unlike A3C, these workers share the same set of policy parameters from the host client and upload their own experiences to the same replay buffer for unified policy updating. %Note that each environment copy is shared by one worker from each individual agent.
The motivation is that the independent randomnesses in different environment copies make the generated experiences mutually independent. This effectively reduces the correlation among the sampled experiences for updating the neural networks, thus improving the learning performance. 

%-------------------------------------------------------
\subsection{Algorithm Design with Enhanced Robustness}
%-------------------------------------------------------
Another challenge lies in the algorithm robustness against the uncertain number of active UAVs. % and the underlying user distributions. 
As UAVs' quit's and join-in's are hardly predictable in the design and training stage, the algorithm needs to handle random quit and join-in given arbitrary number of active UAVs below a maximum. A conceptually simple idea is to train a series of agents to handle quit and join-in separately for different numbers of active UAVs. But that would lead to a series of trainings with prohibitively high training load that increases superlinearly with the maximum number of UAVs. Novel training methods that impose much less training load yet can still accommodate arbitrary changes in UAV crew are much needed. %In addition, the user distributions may differ in different missions and even vary during one mission. Thus in many practical situations, user distributions are hard to know in advance. How to achieve a universal UCN management strategy that is robust to different user distributions is still an open challenging issue.

The most challenging point to achieve this robustness %against the uncertain number of active UAVs 
is minimizing the training load. Obtaining a robust algorithm with one single training is technically feasible, if sufficient experiences concerning every possible change are collected and used to train the agent(s). This requires deliberate episode design to be inclusive yet concise enough while assuring fair observation among all possible changes. Instead of homogeneous episode design, heterogeneous design may be used, where episodes with different situations of change can take place in turns. This simplifies the design of each episode and thus the exploration complexity. %To achieve the robustness against unpredictable user distributions, one potential method is to integrate convolutional neural networks (CNNs) into the DRL design. A CNN structure can be plugged into the agent network to leverage on its strong capability of extracting high-level features from 2D data set. During training, a pool of user distributions can be fed to the agent as part of the states in different episodes. This makes the agent generalize its strategy based on the extracted features. During execution, the UAV crew will first explore the unknown region to gain a user distribution map. The agent(s) will use the map to infer an optimal UCN configuration directly.

%===============================================================================================
\section{Proactive UAV Control Strategy in Solar-Powered Self-Sustainable UCNs}\label{sec.4}
%===============================================================================================
When a solar-powered UAV quits for charging, the quit may lead to failure in satisfying user service demand. But if the UAV is not charged in time, the UCN cannot be sustained. The key out of this dilemma lies in the time-variability of user service demands. When users are spatially concentrated or user service demands are low, fewer UAVs are needed. Some UAVs can be opportunistically scheduled for solar charging even if they are not in immediate need. They can be later called back to meet the increased demand or replace other UAVs. Such proactive control on the serving UAV crew can balance the user service satisfaction and UCN self-sustainability, making UCNs free of charging facility with high energy utilization. This motivates the design of UAV solar-charging strategies, aiming to optimize the target network performance over a time horizon, subject to the constraints of user service demand and network sustainability. 

\subsection{Algorithm Design and Problem Decomposition}
A standard design of the RL algorithms for proactive control will involve a state space including at least the UAV battery residual, the serving status (being idle, serving or charged), user service fulfillment conditions, UAV 3D positions, and time (to help capture environment dynamics). The action space will at least consist of the moving direction(s) and distance(s), and instructions to UAVs on whether to go charging, serving, or idle. The hard constraints will be the UAV sustainability and user service requirements.
%The user requirements can be either individual ones (e.g., throughput) or overall ones (e.g., overall ratio of the served users). 
In addition, the learning needs to consider the environmental dynamics such as dynamic solar radiation intensity and time-varying user distribution and service demand at different times of the day. 
However, such a design will result in a high-dimension state space and a multi-time-scale action space (i.e., movements related actions at a smaller time scale and charging decisions at a larger time scale), making the algorithm prohibitively complex. 

To tackle this issue, one potential way is to decompose the original learning problem into two sub-problems. In the first sub-problem, given the user service demand in each time step, the least number of UAVs and their optimal 3D positions to meet the demand will be determined via Deep RL (DRL). The outputs will be step-specific and combined to form the time-varying constraints for the second sub-problem. In the second sub-problem, given the initial UAV battery residuals and time-varying solar radiation intensity, DRL algorithms can be designed to decide the optimal charging profiles for each UAV, i.e., whether a UAV should go idle, serving or charging in each step, subject to the output of the first sub-problem and the network sustainability. By conducting one DRL training for each time step (the first sub-problem) and a single DRL training over the entire time horizon (the second sub-problem), the original problem can be tractable. 

\subsection{Fusion of Game Theory in MARL Framework with Hybrid Cooperate-Compete Relationship} 
When MARL is leveraged, existing studies on the UCNs usually assume full cooperation among UAVs neglecting UAVs’ own benefits. Yet in many occasions, UAVs may have a hybrid relationship, i.e., cooperate and compete at the same time. Each UAV may want to maximize its own battery residual at the end while working together to satisfy the network constraints. Under such a hybrid relationship, simply following the existing design as described in \ref{sub.design_Elements} may lead to unsatisfying distributed policies\cite{pham2018cooperative}. Novel approaches must be identified or developed to model the cooperative-competitive interaction among UAVs and guide the value function updates for each UAV during training.

To handle such a hybrid relationship, the learning rewards of each UAV need to be revised to include at least two parts: one linked to the overall network objectives, and the other being offset from each other due to the contention for charging and idling opportunities. As the contention among UAVs naturally forms a game, it can be promising to integrate game theory into MARL framework to guide the learning evolution. The result will be a joint coordinated policy that lead to an equilibrium. Two game theory based RL techniques, i.e., Nash Q-learning\cite{casgrain2022deep} and correlated Q-learning\cite{tsai2020achieving}, can be potentially exploited to achieve Nash and Correlated equilibra, respectively. Nash Q-learning requires each UAV agent to maintain a Q-function for its local state space and a joint action space for all UAVs. In each iteration, each agent selects its own action independently, but needs to calculate a joint Nash equilibrium via quadratic programming and update its Q-function based on the Nash Q-function. In correlated Q-learning, each agent maintains a Q-function for the joint state space and the joint action space of all UAVs. In each iteration, different Nash Q-learning, each agent will agree on a joint action selection by solving a linear programming to achieve the correlated equilibrium. Compared to the correlated Q-learning, Nash Q-learning requires each agent to maintain a smaller Q memory but with more computation complexity to calculate the equilibrium. The two techniques can be evaluated in terms of convergence speed, memory and computation complexity, battery residuals, and target network performance.

%===============================================================================================
\section{Case Studies}\label{Sec.CaseStudy}
%==============================================================================================
As illustrative examples, in this section, we introduce how to apply different RL algorithms to regulate UCNs with dynamic UAV set. 

\subsection{Responsive Regulation with Centralized DRL in General UCNs}\label{subsec.responsive_central}
%We first study a ground user service coverage problem in a general UCN\cite{zhang2021learning}. 
As shown in Fig. \ref{fig.systemmodel}, we consider that a group of UAVs flying over a target region to provide communication services to the ground users. A user is served only when its minimum throughput is satisfied. We target at an optimal UAV control policy to maximize the total number of served users over time. The policy responsively relocates the UAVs when \textit{i)} a UAV quits or joins the network, or \textit{ii)} the underlying user distribution changes\cite{zhang2021learning}.

A centralized DRL approach is designed using deep deterministic policy gradient (DDPG) algorithm to accommodate the continuous state and action space. In the state space, all the UAV positions are included. To identify the upcoming change in the UAV set, UAV battery residuals (case of quit) and join-in countdown timer (case of join-in) are included. The time is also included to capture the dynamics of user distribution. The action space includes UAVs' flying directions and distances. The reward awards the total number of served users and punish UAVs for going out of bound. To handle the change in the UAV set, fixed-dimension raw state space and action space are maintained. The raw space corresponds to a maximum possible number of active UAVs that may be involved, but only the states and actions of the active UAVs (i.e., the serving and joining UAVs) will be updated and contribute to the reward. Moreover, to promote the learning exploration, the APC structure in Subsection \ref{subsec.1_promote} is exploited. 

\begin{figure}[!ht]
%\begin{minipage}[b]{0.45\textwidth}
\centering
\includegraphics[width=3.4in, height = 1.6in]{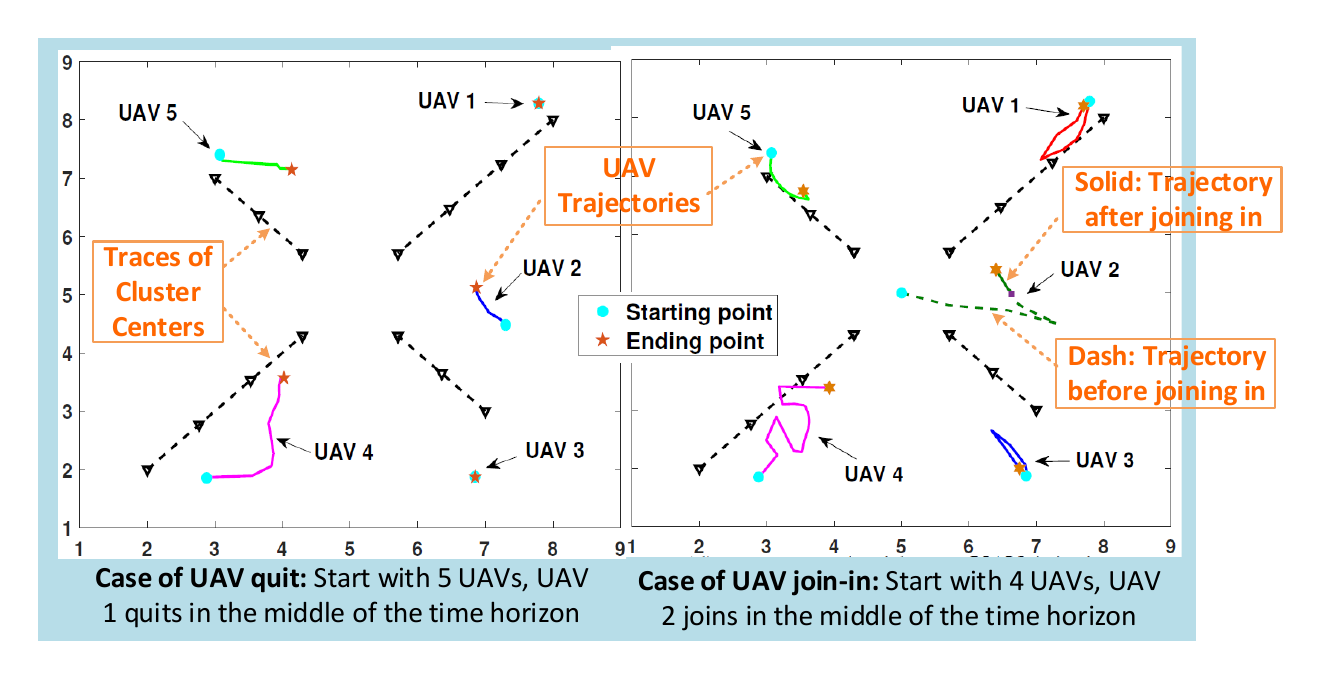}
\caption{\small{UAV trajectories with dynamic user distribution in cases of UAV quit and join-in, respectively.}}
\label{fig.centralized_trajectory}
%\end{minipage}
%\hspace{0.02\textwidth}%
%\begin{minipage}[b]{0.45\textwidth}
%\centering
%\includegraphics[width=3.2in]{results.eps}
%\caption{\footnotesize{Average user throughput comparisons in both licensed and unlicensed spectrum under different deployment patterns.}}
%\label{fig.results}
%\end{minipage}%
\end{figure}

With the above design, the obtained strategy can identify any upcoming change in the UAV set and captures the dynamics of the underlying user distribution. It adaptively relocates the active UAVs when a change happens or even ahead of the change if necessary. Fig. \ref{fig.centralized_trajectory} presents bird views of UAV trajectories obtained by the strategy. The users are mostly distributed in 4 clusters with centers moving towards the region center. When a UAV quits, the remaining UAVs mostly (UAVs 2, 4, 5) move towards the center along with the clusters, but adjust their trajectories towards the quitting UAV (UAV 1) to cover the service hole as much as possible. When a UAV joins in, the existing UAVs (UAVs 1, 3, 4, 5) initially move towards the center, but may turn around (i.e., UAVs 1 and 3) when UAV 2 is integrated. 

\subsection{Distributed Regulation with MARL in General UCNs}\label{subsec.responsive_dis}
This subsection considers the same problem and network setup as Subsection \ref{subsec.responsive_central}, but provides a multi-agent DQL based approach to obtain a distributed regulation strategy\cite{zhang2024distri}. %Each UAV has its own agent that learns to decide its own movements yet in a coordinated way. The decision is based on UAV's local states and information sharing among UAVs. 
In each step, each UAV shares its position, away/active status, and the number of its served users in each step. The individual state space includes its own position, the away/active status of all the UAVs that may be involved, and time. The individual action space only considers its own movements. The reward awards the total number of served users, and punishes the coverage overlapping between this UAV and the others. Such penalty makes the UAVs more dispersed, thus potentially reducing competition and serving more users.

\begin{figure*}[!ht]
  \centering
  \includegraphics[width=6.8in, height=1.6in]{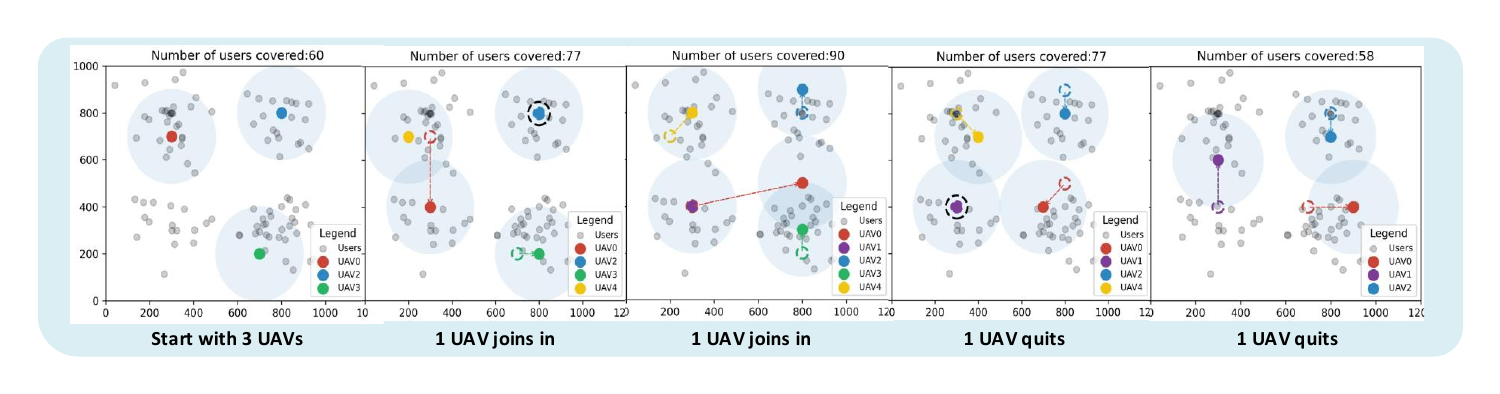}\\
  \caption{\small{Optimal coverage of active UAVs when UAVs randomly quit and join in sequentially.}}\label{fig.distri}
\end{figure*}

A distributed strategy is yielded which can handle arbitrary UAV quit's and join-in's by taking the following innovative measures. In each episode, UAVs randomly and sequentially quit from the maximum possible number to only 1. Two environment copies with the same setting except the randomness are considered. When a UAV quits in one copy, it automatically enters the other copy where its agent continues training with any existing UAVs for the same optimization goals. The UAV sets in the two copies are complementary and their trainings take place simultaneously. The collected experiences of the same UAV from two copies are combined in a single buffer
to update its Q-network. This ensures statistically fair traverse of all possible changes and thus the robustness against the arbitrariness. Fig. \ref{fig.distri} shows bird views of UAV trajectories in one example of random changes in UAV set. The hollow and solid dots represent the optimal UAV positions before and after a change happens, respectively. It can be seen that each time when a random change happens, the obtained strategy is able to relocate the active UAVs to maximize the user coverage.

\subsection{Proactive UAV Control in Solar-Powered UCNs}\label{subsec.proactive}
\begin{figure}[!ht]
  \centering
  \includegraphics[width=3.0in]{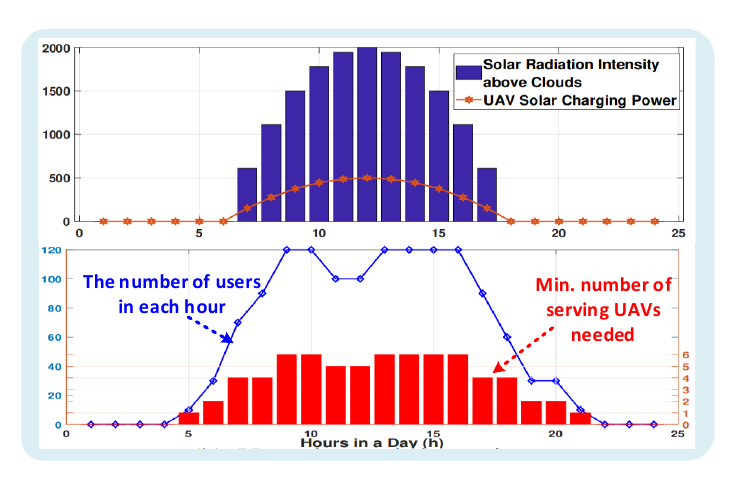}\\
  \caption{\small{Dynamics of solar radiation and user service demand in a day.}}\label{fig.Env_Dynamics}
\end{figure}

With solar-chargeable UAVs, solar charging strategies can be designed to optimize when each UAV goes serving, charging or idle (i.e., to the ground for energy saving). The change of the serving UAV set can then be proactively controlled for better network management as compared to reacting passively to uncontrollable changes in general UCNs (i.e., Subsections \ref{subsec.responsive_central} and \ref{subsec.responsive_dis}). To this end, our work \cite{wang2023optimal} studies UAV solar charging profile design, aiming to balance between maximizing the total number of served users over time and maximizing the total UAV residual energy at the end. In each step, the design needs to meet the minimum overall user service percentage and the UCN sustainability requirements, i.e., every UAV needs to have sufficient energy to go charging at the end of each step. The design also considers time-varying solar radiation and user service demand at a day scale (Fig. \ref{fig.Env_Dynamics}), which makes the problem more realistic but challenging.
%\vspace{-4mm}

\begin{figure}[!ht]
  \centering
  \includegraphics[width=3.0in]{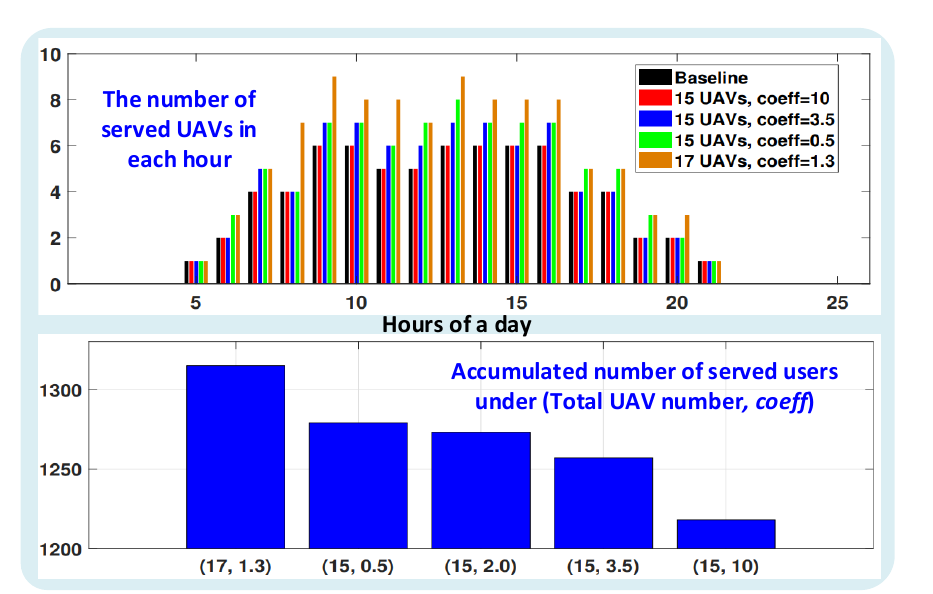}\\
  \caption{\small{Performance demonstration under different parameters.}}\label{fig.Solarcharging}
\end{figure}

We decouple the original problem into two sub-problems. Sub-problem 1 calculates mappings between the number of serving UAVs and the maximum served users given hourly user service demands. It can be resolved by reusing the algorithms designed in Subsection \ref{subsec.responsive_central}. Sub-problem 2 designs UAV charging profiles based on the obtained mappings, solar radiation dynamics, UAV sustainability, and the minimum overall user service percentage. A centralized DDPG algorithm is designed to solve the problem, with a relaxation mechanism to handle the large discrete action space. Simulation results are shown in Fig. \ref{fig.Solarcharging}. Whole sets of 15 and 17 UAVs are simulated, respectively, with different \textit{coeff} values. A smaller \textit{coeff} favors serving more users over saving more UAV energy. The baseline is the minimum number of UAVs needed in each hour to meet the minimum overall user service percentage. With smaller \textit{coeff}, more UAVs tend to be scheduled to serve, with less chance of being charged. This leads to more served users over time but less residual energy at the end. 

%===============================================================================================
\section{Conclusions and Future Outlook}\label{sec.Conclusion}
%===============================================================================================
In this article, we have discussed how to adaptively regulate UCNs given a dynamic set of UAVs under the RL framework. System overview, research directions, design challenges, potential approaches and case studies have been provided from aspects of reactive strategies in general UCNs and proactive strategies in solar-powered UCNs. Illustrative examples have demonstrated that deep RL algorithms can well handle arbitrary changes in UAV set reactively and control the change of the serving UAV set proactively via UAV solar-charging profile design. However, there are still many open issues to be addressed and opportunities to be explored, some of which are highlighted below. 
\subsection{Open Issues}
\textbf{Robustness against unknown user distribution}. The existing works are mostly based on known or predicted user distributions. Robust regulation strategies are desired to handle random user distributions for unexplored environment. A potential solution is to integrate Convolutional Neural Networks (CNNs) into the RL framework, and analyze the live user distribution map as part of the state space. 

\textbf{Hybrid relationship among UAVs}. When designing the UAV solar-charging profiles, a practical consideration is that UAVs may contend against each other (e.g., for solar charging opportunities) while cooperating to meet overall objectives. This hybrid relationship may make the existing RL algorithms perform unsatisfactorily. Game theories may be introduced to guide the learning to a better convergence by coordinating UAVs' action decisions in each step for equilibrium.

\subsection{Opportunities of Broader Scope}
\textbf{Integration of generative AI (GAI)}. GAI has achieved great success in content creation. Its remarkable capability of complex task processing can be exploited in regulating UCNs with dynamic UAV set. For instance, the large language models (LLMs) can be utilized to make accurate time-series prediction of the environment and UAV dynamics. Generative Adversarial Networks (GANs) can be adopted to produce synthetic experiences that resemble the realistic ones. This may achieve learning expedition and better generalization to different change cases of the UAV set.   

\textbf{Wireless charging of UAVs}. In addition to solar charging, wireless UAV charging is another innovative technology for which proactive control of the UAV set can be investigated. There are several forms of wireless charged UCNs. Ground charging docks or towers can be deployed such that UAVs may quit the network to get charged while being landed (docks) or on the fly (towers). Airships can also be exploited like tanker planes so that UAVs can get charged over the air without leaving the network. Laser beaming is another option to charge over the air but may be limited by the locations.

%%%%%%%%%%%%%%%%%%%%%%%%%%%%%%%%%%%%%%%%%%%%%%%%%%%%%%%%%%%%%%%%%%%%%%%%%%%%%%%%%%%%%%%%%%%%%%%%%%%%%%%%%%%%%%
\bibliographystyle{IEEEtran}
\bibliography{reference}

\section*{Acknowledgements}
This work is supported by Natural Science Foundation under Award 2412393.

\section*{Biography}
%\begin{description}
%\item 
\small{\textbf{Ran Zhang} (rzhang8@charlott.edu) [SM'22] is an Assistant Professor with Department of Electrical and Computer Engineering, University of North Carolina at Charlotte, USA.}

\small{\textbf{Bowei Li} (boweili@andrew.cmu.edu) is a Master student with Department of Electrical and Computer Engineering, Carnegie Mellon University, USA.}

\small{\textbf{Liyuan Zhang} (lzhang51@charlotte.edu) is a PhD student with Department of Electrical and Computer Engineering, University of North Carolina at Charlotte, USA.}

\small{\textbf{Jiang (Linda) Xie} (jxie1@charlotte.edu) [F'20] is a Professor with Department of Electrical and Computer Engineering, University of North Carolina at Charlotte, USA.}

\small{\textbf{Miao Wang} (mwang25@charlotte.edu) [SM'21] is an Assistant Professor with Department of Electrical and Computer Engineering, University of North Carolina at Charlotte, USA.}

%\end{description}

\end{document}